\documentclass[11pt]{article}
\usepackage{graphicx}  
\usepackage[final]{epsfig}
\usepackage{amssymb,latexsym}
\usepackage[mathscr]{eucal}
\usepackage{citesort}
\usepackage{url}
\usepackage{ntheorem}

\newcommand{\ii}{\mathrm{i}}

\newcommand{\zf}{\mathring f}
\newcommand{\zb}{\mathring b}
\newcommand{\bz}{\mathring b}

\newcommand{\hE}E
\newcommand{\hR}R
\newcommand{\hm}m

\newcommand{\FS}       
                  {F}

\newcommand{\HS} 
       {H_{\mbox{\scriptsize volume}}}

{\ptc{this should be removed in the oberwolfach version}}%

\newcommand{\mcE}{{\mycal E}}%

\newcommand{\eeal}[1]{\label{#1}\end{eqnarray}}
\newcommand{\C}{{\mathbb C}}
\newcommand{\bed}{\begin{deqarr}}
\newcommand{\eed}{\end{deqarr}}
\newcommand{\bedl}[1]{\begin{deqarr}\label{#1}}
\newcommand{\eedl}[2]{\arrlabel{#1}\label{#2}\end{deqarr}}

\newcommand{\loc}{\textrm{\scriptsize\upshape loc}}

\newcommand{\bel}[1]{\begin{equation}\label{#1}}
\newcommand{\bea}{\begin{eqnarray}}
\newcommand{\bean}{\begin{eqnarray}\nonumber}
\newcommand{\beal}[1]{\begin{eqnarray}\label{#1}}
\newcommand{\eea}{\end{eqnarray}}

\newcommand{\Eq}[1]{Equation~\eq{#1}}

\newcommand{\qed}{\hfill $\Box$ \medskip}
\newcommand{\proof}{\noindent {\sc Proof:\ }}
\newcommand{\be}{\begin{equation}}
\newcommand{\eeq}{\end{equation}}
\newcommand{\ee}{\end{equation}}
\newcommand{\beqa}{\begin{eqnarray}}
\newcommand{\eeqa}{\end{eqnarray}}
\newcommand{\beqan}{\begin{eqnarray*}}
\newcommand{\eeqan}{\end{eqnarray*}}
\newcommand{\ba}{\begin{array}}
\newcommand{\ea}{\end{array}}

\newcommand{\mcM}{{\mycal M}}

\newtheorem{Theorem} {\sc  Theorem\rm} [section]

\newtheorem{Lemma} [Theorem] {\sc  Lemma\rm}
\newtheorem{Proposition} [Theorem] {\sc  Proposition\rm}

\theorembodyfont{\upshape}
\newtheorem{Remark}[Theorem]{\sc  Remark\rm}

\DeclareFontFamily{OT1}{rsfs}{}
\DeclareFontShape{OT1}{rsfs}{m}{n}{ <-7> rsfs5 <7-10> rsfs7 <10-> rsfs10}{}
\DeclareMathAlphabet{\mycal}{OT1}{rsfs}{m}{n}

{\catcode `\@=11 \global\let\AddToReset=\@addtoreset}
\AddToReset{equation}{section}

\newcounter{mnotecount}[section]

\renewcommand{\themnotecount}{\thesection.\arabic{mnotecount}}

\newcommand{\mnote}[1]
{\protect{\stepcounter{mnotecount}}$^{\mbox{\footnotesize
$
\bullet$\protect\themnotecount}}$ \marginpar{
\raggedright\tiny\em
$\!\!\!\!\!\!\,\bullet$\protect\themnotecount: #1} }

\newcommand{\warn}[1]
{\protect{\stepcounter{mnotecount}}$^{\mbox{\footnotesize
$
\bullet$\themnotecount}}$ \marginpar{
\raggedright\tiny\em
$\!\!\!\!\!\!\,\bullet$\protect\themnotecount: {\bf Warning:} #1} }

\newcommand{\R}{\mathbb R}

\newcommand{\N}{\mathbb N}

\newcommand{\eq}[1]{(\ref{#1})}

\newcommand{\ptc}[1]{\mnote{{\bf ptc:}#1}}

\newcommand{\beqar}{\begin{deqarr}}
\newcommand{\eeqar}{\end{deqarr}}

\newcommand{\beaa}{\begin{eqnarray*}}
\newcommand{\eeaa}{\end{eqnarray*}}

\newcommand{\zg}{\mathring{g}}

\begin{document}

\title{The Ernst equation and ergosurfaces}

\author{Piotr T.\ Chru\'sciel\thanks{Current address: A.~Einstein Institute, Golm. Partially supported by a Polish
Research Committee grant 2 P03B 073 24. E-mail
\protect\url{Piotr.Chrusciel@lmpt.univ-tours.fr}, URL \protect\url{
www.phys.univ-tours.fr/}$\sim$\protect\url{piotr}} \\ LMPT,
F\'ed\'eration Denis Poisson,
Tours
 \\
  \\Gert-Martin Greuel\thanks{{ E--mail}:
  \protect\url{greuel@mathematik.uni-kl.de}} \\ Universit\"at Kaiserslautern
 \\ \\ Reinhard Meinel\thanks{Partially supported by the Deutsche Forschungsgemeinschaft SFB/TR7-B1.{ E--mail}:
  \protect\url{meinel@tpi.uni-jena.de}} \\ Universit\"at Jena
\\
  \\ Sebastian J. Szybka\thanks{Partially supported by a Polish
Research Committee grant 1 P03B 012 29 { E--mail}:
  \protect\url{szybka@if.uj.edu.pl}} \\ Obserwatorium Astronomiczne UJ,
Krak\'ow}

\maketitle
\begin{abstract}
  We show that analytic solutions $\mcE$ of the Ernst equation with
  non-empty
  zero-level-set of $\Re \mcE$ lead to smooth ergosurfaces in space-time.
  In fact, the space-time metric is smooth near a ``Ernst ergosurface" $E_f$ if
  and only if
  $\mcE$ is smooth near $E_f$ and does not have
  zeros of infinite order there.
\end{abstract}

\newcommand{\beq}{\begin{equation}}

\newcommand{\zX}{\mathring{X}}%

\section{Introduction}
 \label{SI}

A standard procedure for constructing stationary axi-symmetric
solutions of the Einstein equations proceeds by a reduction of the
Einstein equations to a two-dimensional nonlinear equation --- the
Ernst equation \cite{Ernst} --- using the asymptotically timelike
Killing vector field $X$ as the starting point of the reduction:
One finds a complex valued field
\begin{equation}\label{2.11a}
\mcE=f+\ii b, \ee
 by e.g. solving a
boundary-value problem~\cite{Neugebauer:2003qe}. The space-time
metric is then obtained by solving ODEs for the metric functions.
The following difficulties arise in this construction:
\begin{enumerate}
\item Singularities of $\mcE$, which might --- or might not --- lead
to singularities of the metric.

\item Struts or causality violations at the rotation axis.

\item Singularities of the equations arising at zeros of $\Re
\mcE$.
\end{enumerate}

The aim of this work is to address this last question. Indeed, the
equations determining the  metric functions are singular at the
zero-level-set\footnote{The equations are of course singular at
$f=\rho=0$ as well, but the singularity of the whole system of
equations has a different nature there, because of the
$\partial_\rho f/\rho$ terms in the Ernst equation \eq{2.11}, and
will not be considered here. Geometrically, the set $\{\rho=f=0\}$
has a rather different nature, corresponding to Killing horizons,
with the boundary conditions there being reasonably well understood
in any case~\cite{LP1,CarterlesHouches}.}
$$E_f:=\{f=0\;,\ \rho>0\}$$
of $f:=\Re \mcE$;  we will refer to $E_f$ as the
\emph{$\mcE$-ergosurface}. We show, assuming smoothness of $\mcE$
in a neighborhood of $E_f$, and excluding zeros of infinite order,
that the singularities of the solutions of those ODEs conspire to
produce a smooth space-time metric. More precisely, we have:

\begin{Theorem}
\label{TA}  Consider a smooth solution $f+\ii b$ of the Ernst
equations \eq{2.11na}-\eq{2.11nb} below such that $f$ has
\underline{no zeros of infinite order} at the $\mcE$--ergosurface
$E_f$. Then there exists a neighborhood of $E_f$ on which the
metric \eq{2.1} obtained by solving \eq{2.12}-\eq{2.13} is smooth
and has Lorentzian signature.
\end{Theorem}

An immediate corollary of Theorem~\ref{TA} is the following: Any
point $\vec x_0$ off the axis in the Weyl coordinate chart
corresponds to  space-time points at which the metric is regular
$\Longleftrightarrow$ the Ernst
 potential is a real-analytic function of the Weyl coordinates near $\vec x_0$
$\Longleftrightarrow$ the Ernst
 potential is a smooth function of the Weyl coordinates near $\vec x_0$ and zeros of $\Re \mcE$ have finite
order.

The proof of Theorem~\ref{TA} can be found at the end of
Section~\ref{Szos}.

 The condition of zeros of finite order is
necessary, in the following sense: any zero of $\Re \mcE$ on a
smooth space-time ergosurface is of finite order. This is proved
at the end of Section~\ref{Sfee}.

It is an interesting consequence of our analysis below that a
critical zero of $f$ of order $k$  corresponds to a smooth
two-dimensional surface in space-time at which  $k$ distinct
components of the ergoregion $\{f<0\}$ ``almost meet", in the sense
that their closures  intersect there, with the boundaries branching
out. Two exact solutions with this behavior for $k=2$ are presented
in Section~\ref{Sncz}.

Section~\ref{STm} below appeared in preprint form in~\cite{CMSNI};
the reader will also find in~\cite{CMSNI} some more information
about second order zeros of $f$.

\section{The field equations and ergosurfaces}
\label{Sfee}

 We consider a vacuum gravitational field in
Weyl-Lewis-Papapetrou coordinates
\bel{2.1}
ds^2=f^{-1}\left[h\left(d\rho^2+d\zeta^2\right)+\rho^2d\phi^2\right]-f\left(d
t+ad\phi\right)^2 \ee
 with all functions depending only upon $\rho$ and $\zeta$.  The
vacuum Einstein equations for the metric functions $h$, $f$, and $a$
are equivalent to the Ernst equation
\bel{2.11}
(\Re \mcE)\left(\mcE_{,\rho\rho}+\mcE _{,\zeta\zeta}+\frac{1}{\rho}\mcE _{,\rho}\right)=\mcE _{,\rho}^2 + \mcE _{,\zeta}^2
\ee
for the complex function $\mcE(\rho,\zeta)=f(\rho,\zeta)+\ii
b(\rho,\zeta)$, where $b$ replaces $a$ via
\bel{2.12}
a_{,\rho}=\rho f^{-2}b_{,\zeta},\quad a_{,\zeta}=-\rho f^{-2}b_{,\rho}
\ee
and $h$ can be calculated from
\bel{2.13}
h_{,\rho}=\frac{\rho
h}{2f^{2}}\left[f_{,\rho}^2-f_{,\zeta}^2+b_{,\rho}^2
-b_{,\zeta}^2\right],\quad h_{,\zeta}=\frac{\rho
h}{f^{2}}\left[f_{,\rho}f_{,\zeta}+ b_{,\rho}b_{,\zeta}\right].
 \ee
%
 We will
think of $\rho$ and $\zeta$ as being cylindrical coordinates in
$\R^3$ equipped with the flat metric
$$
\zg=
d\rho^2+\rho^2d\varphi^2+d\zeta^2\;,
$$
with all the above functions being $\varphi$--independent functions on
$\R^3$. Then \eq{2.11} can be rewritten as
\beal{2.11na}
&
f \Delta f = |Df|^2-|Db|^2\;,
&
\\
&
f \Delta b = 2 (Df,Db)\;.
&
\eeal{2.11nb}
where $\Delta$ is the flat Laplace operator of the metric $\zg$,
and $(\cdot,\cdot)$ denotes the $\zg$-scalar product, similarly
the norm $|\cdot |$ is the one associated with $\zg$.

Equations \eq{2.11na}-\eq{2.11nb} degenerate at $\{f=0\}$, and it
is not clear that $f$ or $b$ will smoothly extend across
$\{f=0\}$, if at all. In Section~\ref{Sns} below we give examples
of solutions which do \emph{not}. On the other hand, there are
large classes of solutions which are smooth across $\mcE_f$.
Examples can be obtained as follows:

First, every space-time obtained from an Ernst map $\mcE'$
associated to the reduction that uses the axial Killing vector
$\partial_\varphi$ (see, e.g.,~\cite{CarterlesHouches,Weinstein1})
will lead to a solution $\mcE$ as considered here that extends
smoothly across the \emph{space-time ergosurfaces} (if any); recall
that an ergosurface is defined to be a \emph{timelike} hypersurface
where the Killing vector $X$, which asymptotes a time translation in
the asymptotic region, becomes null. Those ergosurfaces correspond
then to $\mcE$-ergosurfaces across which $f$ does indeed extend
smoothly. We emphasise that we are interested in the construction of
a space-time starting from $\mcE$, and we have no \emph{a priori}
reason to expect that an $\mcE$--ergosurface, defined as smooth
zero-level set of $\Re \mcE$, will lead to a smooth space-time
ergosurface.

Next, large classes of further examples are given
in~\cite{KramerNeugebauerDoubleKerr,Neugebauer:2003qe,%
MankoRuiz1,Meinel:Neugebauer:Kleinwaechter,Yamazaki,%
Neugebauer:integral,NeugebauerMeinelPRL}%
\footnote{The solutions we are referring to here are not
necessarily vacuum everywhere, and some of them have a function
$\mcE$ which is singular somewhere in the $(\rho,\zeta)$ plane.
Our analysis applies to the vacuum region, away from the rotation
axis, and away from the singularities of the Ernst map $f+\ii
b$.}.  Some of the solutions in those references have non-trivial
zero-level sets of $\Re \mcE$, with $g_{\rho\rho}=g_{zz}$ and
$g_{t\varphi}$ smooth across $E_f$ (see in particular
\cite{MankoRuiz1}), but smoothness of $g_{\varphi\varphi}$ is not
manifest.

 Quite generally, we
have the following: consider a vacuum space-time $(\mcM,g)$ with
two Killing vectors $X$, $Y$, and  with a non-empty
\emph{space-time ergosurface} defined as
$$
 E_\mcM:=\{\underbrace{g(X,X)=0\;,\ X\ne 0}_{(1)}\;,\
\underbrace{g(X,X)g(Y,Y)-g(X,Y)^2< 0}_{(2)} \}
 \;.
$$
Condition (1)  is the statement that $X$ becomes null on $E_\mcM$,
while (2) says that the planes spanned by $X$ and $Y$ are timelike;
condition (2)  distinguishes a space-time ergosurface from a
horizon, where those plane are null. (For solutions in Weyl form,
condition (2) translates into the requirement  $\rho \ne 0$.) Now,
by (2) there exists a linear combination of $X$ and $Y$ which is
timelike near $E_\mcM$, and if $g$ is sufficiently differentiable
($H^2_\loc$ in coordinates adapted to the symmetry group is more
than enough), the analysis of \cite{MzH} shows that there exist an
atlas near $E_\mcM$ in which $g$ is analytic. By chasing through the
construction of Weyl coordinates, this implies that $f$ and $b$  are
real-analytic functions near $E_f$. In particular $f$ will
\emph{not} have zeros of infinite order there.

\section{Static solutions}
\label{Sns}

In the remainder of this work only those solutions which are
invariant under rotations around some fixed chosen axis are
considered (when viewed as functions on subsets of $\R^3$), and
all functions are identified with functions of two variables,
$\rho$ and $\zeta$.

 Consider a
solution of \eq{2.11na}-\eq{2.11nb} with $b\equiv 0$.  Setting $u
=\ln f$ in the region $\Omega:=\{f>0\}$, \Eq{2.11na} becomes
\bel{ueq}
\Delta u = 0 \ \mbox{ on } \ \Omega \;.
 \ee
One can now obtain examples of solutions for which $E_f$ is not
empty as follows: Let $\alpha \in \R^*$, $\vec
x_0=(\rho_0,\zeta_0)\in (0,\infty)\times \R$; standard PDE
considerations show existence of solutions of \eq{ueq} on
$\Omega:=\{(0,\infty)\times \R\}\setminus \{\vec x_0\}$ such that
$$
u_\alpha = \alpha \ln\Big((\rho-\rho_0)^2+(\zeta-\zeta_0)^2\Big) +
O(1) \;.
$$
This leads to
$$
f_\alpha := e^{u_\alpha} =
\Big((\rho-\rho_0)^2+(\zeta-\zeta_0)^2\Big)^\alpha g_\alpha \;,
$$
where $g_\alpha$ has no zeros. We have the following:
\begin{itemize}
\item No such solution is smoothly extendable through the Ernst
ergosurface $E_{f_\alpha}=\{\vec x_0\}$  except perhaps when $\alpha
\in \N^*$.
\item In that last case the solutions do not extend smoothly
across $E_f$ either, which can be seen as follows. Consider, first
$\alpha=1$, then $f=f_1$ has a zero of order two with
positive-definite Hessian, but Lemma~\ref{Lnosec} below shows that
no such solutions which are smooth across $E_{f}$ exist. For general
$\alpha=n\in \N^*$ we note that
$$
f_{1}=\Big((\rho-\rho_0)^2+(\zeta-\zeta_0)^2\Big) (g_{n})^{\frac 1
{n}} \;.
$$
But smoothness of $f_{n}$ would imply that of $g_{n}$, and thus of
$f_{1}$, which is not the case. Thus $f_{n}$, $n\in \N^*$,  cannot
be smooth either.
\end{itemize}
Above we have considered differentiability of $f_\alpha$ in
$(\rho,\zeta)$--coordinates. This might \emph{not} be equivalent
to the question which is of main interest here, that of regularity
of the space-time metric. In the case $b\equiv 0$ this issue is
easy to handle, by noting that $a$ can then always be made to
vanish by a redefinition of $t$. Now, the length
$g(\partial_\varphi,\partial_\varphi)$ of the Killing vector
$\partial_\varphi$, generating rotations around the axis, is a
smooth  --- hence locally bounded --- function on the space-time.
But $g_{\varphi\varphi}=\rho^2 f^{-1}$ by \eq{2.1} so, in the
static case, zeros of $f$ with $\rho_0\ne 0$  cannot correspond to
ergosurfaces in space-time\footnote{The discussion here gives a
simple proof, under the supplementary condition of axi-symmetry,
of the Vishweshwara-Carter lemma, that there are no ergosurfaces
in static space-times.}.

\section{Non-critical zeros of $f$}
\label{STm}

 We start with the following:

\begin{Theorem}
\label{Tmain}
  The conclusion of Theorem~\ref{TA} holds if one
moreover assumes that $|Df|$ has no zeros at the
$\mcE$--ergosurface $E_f:=\{f=0, \rho>0\}$.
\end{Theorem}

\proof We need to show that the functions
$$
\alpha:=-g_{\varphi t} =af\;,\quad \beta:= \ln g_{\zeta \zeta}=\ln
g_{\rho\rho}=\ln (hf^{-1})\;,
$$
as well as
$$
g_{\varphi\varphi}= \frac{\rho^2 - (af)^2}f
$$
are smooth across $\{f=0, \rho>0\}$, and that $g_{\varphi t}$ does
\emph{not} vanish whenever $g_{tt}=-f$ does.

We start by Taylor-expanding $f$ and $b$ to order two near any point
$(\rho_0,\zeta_0)$ such that $f(\rho_0,\zeta_0)=0$:
\begin{eqnarray*}
f(\rho,\zeta)&=&\zf_{,\rho} (\rho-\rho_0)+\zf_{,\zeta} (\zeta-\zeta_0)
\nonumber
\\
&&
+\frac{1}{2}\zf_{,\rho\rho}
(\rho-\rho_0)^2+\frac{1}{2}\zf_{,\zeta\zeta} (\zeta-\zeta_0)^2+\zf_{,\rho\zeta}
(\rho-\rho_0)(\zeta-\zeta_0) + \dots,
\\
b(\rho,\zeta)&=&\zb +\bz_{,\rho}
(\rho-\rho_0)+\bz_{,\zeta} (\zeta-\zeta_0)
\\
&&
+\frac{1}{2} \bz_{,\rho\rho}
(\rho-\rho_0)^2+\frac{1}{2}\bz_{,\zeta\zeta} (\zeta-\zeta_0)^2+\bz_{,\rho\zeta}
(\rho-\rho_0)(\zeta-\zeta_0) + \dots,
\end{eqnarray*}
where a circle over a function indicates that the value at $\rho_0$
and $\zeta_0$ is taken. Inserting these expansions into
\eq{2.11na}-\eq{2.11nb}, after tedious but elementary algebra one
obtains either
\begin{equation}
\begin{array}{cccc}
&\bz_{\rho} =\eta \zf_{\zeta} ,&\bz_{\zeta} =-\eta \zf_{\rho} ,
\quad \eta=\pm 1
&\\
\zf_{,\rho\rho} +\zf_{,\zeta\zeta} =
\frac{\zf_{,\rho} }{\rho_0}, & \bz_{,\rho\rho} +\bz_{,\zeta\zeta}
=\frac{\zf_{,\zeta} }{\rho_0}, & \bz_{,\rho\zeta} =\zf_{,\zeta\zeta} ,
& \zf_{,\rho\zeta} =\bz_{,\rho\rho} ,\\
\end{array}
\label{3.1}
\end{equation}
or
\begin{equation}
\bz_{\rho} = \zf_{\zeta} =\bz_{\zeta} = \zf_{\rho}=0\;.
\end{equation}
The second possibility is excluded by our hypothesis that $Df\ne 0$ on
$E_f$.

Suppose, first, that $\eta=1$ in the first line of \eq{3.1}. From
\eq{2.12} we obtain
\beal{eq1}
\alpha_{,\rho} & = & \frac {f_{,\rho}}f \alpha + \frac \rho f
b_{,\zeta}\;,
 \\
 \alpha_{,\zeta} & = & \frac {f_{,\zeta}}f \alpha - \frac \rho f
b_{,\rho} \;,
 \eeal{eq2}
so that
\beal{eq1n}
\left(\frac{\alpha-\rho}f\right)_{,\rho}
& = & \underbrace{[\rho(b_{,\zeta}+f_{,\rho})-f]}_{=:\sigma_\rho}f^{-2}
\;,
 \\
\left(\frac{\alpha-\rho}f\right)_{,\zeta} & = & \underbrace{\rho(f_{,\zeta}-b_{,\rho})}_{=:\sigma_\zeta}
f^{-2}
\;.
 \eeal{eq2n}
Inserting \eq{3.1} into the definitions of $\sigma_\rho$ and $\sigma_\zeta$ we find
$$
\sigma_\rho=\sigma_\zeta=0=d\sigma_\rho=d\sigma_\zeta
$$
at every point $(\rho_0,\zeta_0)$ lying on the $\mcE$-ergosurface.
Here, as elsewhere, $d\mu$ denotes the differential of a function
$\mu$.

Recall  that $Df$ does not vanish on $E_f$. We can thus introduce
coordinates $(x,y)$ near each connected component of $E_f$ so that
$f=x$. Since the $\sigma_a$'s are smooth we have the Taylor
expansions
$$
\sigma_a= \sigma_a|_{E_f}+ (\partial_x \sigma_a)|_{E_f}x +
 r_ax^2 \;,
$$
for some remainder terms $r_a$ which are smooth functions on space-time.
But we have shown that $ \sigma_a|_{E_f}= (\partial_x \sigma_a)|_{E_f}=0$. Hence
$$
\sigma_a=
 r_ax^2 =r_af^2  \;,
$$
 It follows that the right-hand-sides of
\eq{eq1n}-\eq{eq2n} extend by continuity across $E_f$ to smooth
functions. Hence the derivatives of $(\alpha-\rho)/f$ extend by
continuity to smooth functions, and by integration
\bel{alpha}
\alpha-\rho=f\hat \alpha\;, \ee
for some smooth function $\hat \alpha(\rho,\zeta)$. This proves
smoothness both of $g_{t\varphi}$ and of $g_{\varphi\varphi}$. We
also obtain that $g_{t\varphi}=-\rho$ when $f=0$, and since $\rho>0$
by assumption we obtain non-vanishing of $g_{t \varphi}$ on  the
$\mcE$--ergosurface.

In the case where $\eta=-1$ in \eq{3.1}, instead of
\eq{eq1n}-\eq{eq2n} we write equations for $(\alpha+\rho)/f$, and an
identical argument applies.

We pass now to the analysis of $g_{\rho\rho}=g_{zz}$. From \eq{2.13},
\beal{2.13n}
\ln (h/f)_{,\rho}& = &
\underbrace{\frac{1}{2}\left[\rho(f_{,\rho}^2-f_{,\zeta}^2+b_{,\rho}^2
-b_{,\zeta}^2)-2ff_{,\rho}\right]}_{=:\kappa_\rho}f^{-2},
\\
\ln (h/f)_{,\zeta} &=&\underbrace{\left[\rho(f_{,\rho}f_{,\zeta}+
b_{,\rho}b_{,\zeta})-ff_{,\zeta}\right]}_{=:\kappa_\zeta}f^{-2}.
 \eeal{2.13n2}
Evaluating $\kappa_a$ and its
derivatives on $E_f$ and using \eq{3.1} one obtains again
$$\kappa_a=d\kappa_a=0$$ on $E_f$.  As before we conclude that
$g_{\rho\rho}$ and $g_{\zeta\zeta}$ are smooth across $E_f$.

\qed

\section{Higher order zeros of $f$}
\label{Sncz}

We shall say that $f$ has a zero  of order $n$, $n\ge 1$, at $\vec
x_0=(\rho_0,\zeta_0)$, if
$$f(\vec x_0)=\ldots =\underbrace{D\cdots D}_{n-1\
\mbox{\scriptsize factors}}f(\vec x_0)=0  \ \mbox{ \underline{but}
} \ \underbrace{D\cdots D}_{n\  \mbox{\scriptsize factors}}f(\vec
x_0)\ne 0 \;.
$$
\newcommand{\text}{\mathrm}
It is legitimate to raise the question whether solutions of the
Ernst equations \eq{2.11} with higher order zeros on $E_f$ exist. A
simple mechanism\footnote{We are grateful to M.~Ansorg and
D.~Petroff for pointing this out to us.} for producing such
solutions is the following: consider a family of solutions depending
continuously on one parameter, such that for large parameter values
there exist two disjoint ergosurfaces, while for small parameter
values the ergosurface is connected. Elementary considerations show
that there exists a value of the parameter where $f$ has a zero of
higher order. Examples of such behavior have been found numerically
by Ansorg~\cite{Ansorg} in families of differentially rotating disks
(however, the merging of the ergosurfaces in that work occurs in the
matter region, which is not covered by our analysis).
 In Figure~\ref{AnsPet}, due to D.~Petroff\footnote{We are very grateful
 to D.~Petroff for providing this figure; a detailed analysis  of
 configurations of this type can be found in~\cite{AnsorgPetroff}.},
 the reader will find an example in a family of solutions with a black hole
 surrounded by a ring of fluid. Those solutions are globally regular, and the coalescence of
 ergosurfaces
 takes place in the vacuum region.
 \begin{figure}[t!]
 \begin{center}
\includegraphics[width=\textwidth]{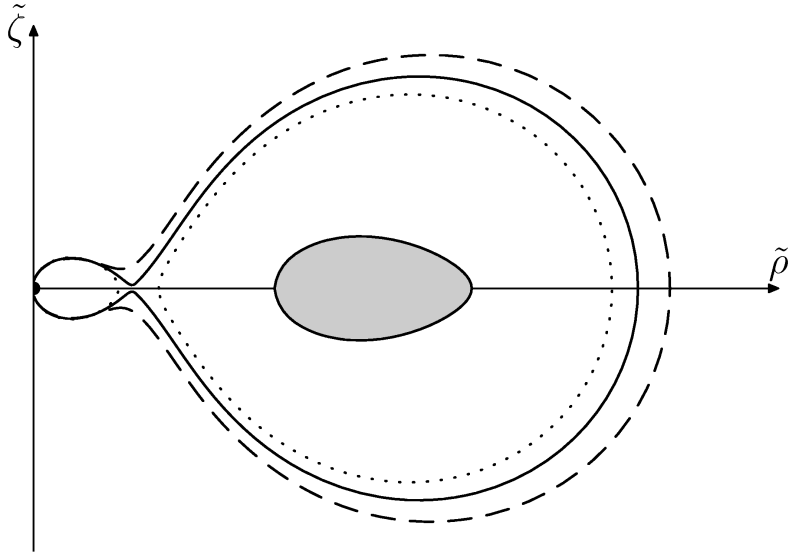}
\end{center}
 \caption[]{A coordinate representation of ergosurfaces
for three configurations consisting of a black hole (coordinate
origin) surrounded by a fluid ring (shaded area). The values of the
parameters, in the notation of~\cite{AnsorgPetroff}, are
$\varrho_{\text i}/\varrho_{\text o}=0.55$, $r_{\text
c}/\varrho_{\text o}=0.015$, $J_{\text c}/\varrho_{\text o}^2=0.05$,
with $V_0=-1.45$ (dashed line), $V_0=-1.396$ (solid line) and
$V_0=-1.35$ (dotted line). The shape of the ring (here corresponding
to the ergosurface indicated by a solid line) is represented by the
shaded area and differs only minimally between the three
configurations. The coordinates $\tilde{\varrho}$ and
$\tilde{\zeta}$ are related to $\varrho$ and $\zeta$ by a conformal
transformation.} \label{AnsPet}
\end{figure}
A purely vacuum example of this kind is hinted at
in~\cite[Fig.~2]{OoharaSato}.
 Finally, Figures~\ref{Coalesc}-\ref{Coalesc2} show a
purely vacuum example within the Kramer-Neugebauer family of
solutions~\cite{KramerNeugebauerDoubleKerr}, where the parameters
which are being varied are the $\beta_i$'s of~\cite{Rueda:2005km}.
\begin{figure}\begin{center}
\includegraphics[height=2in,width=.3\textwidth]{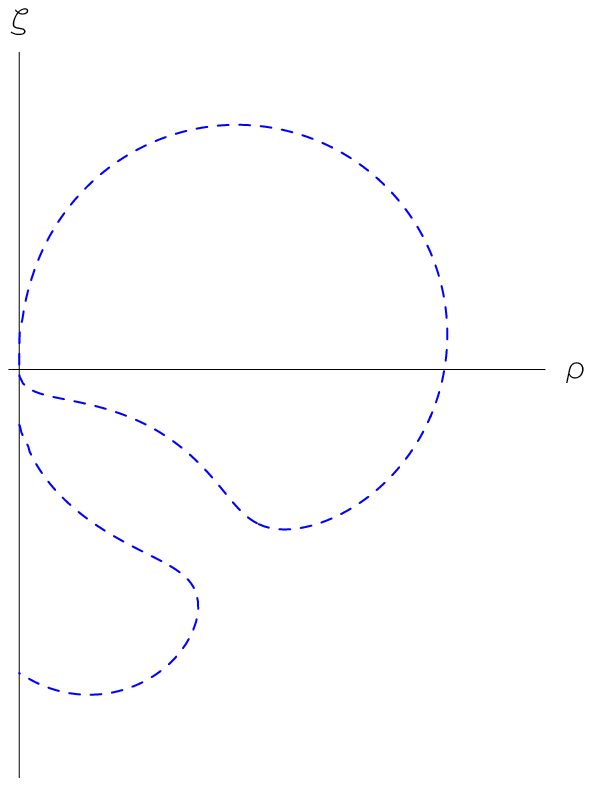}
\includegraphics[height=2in,width=.3\textwidth]{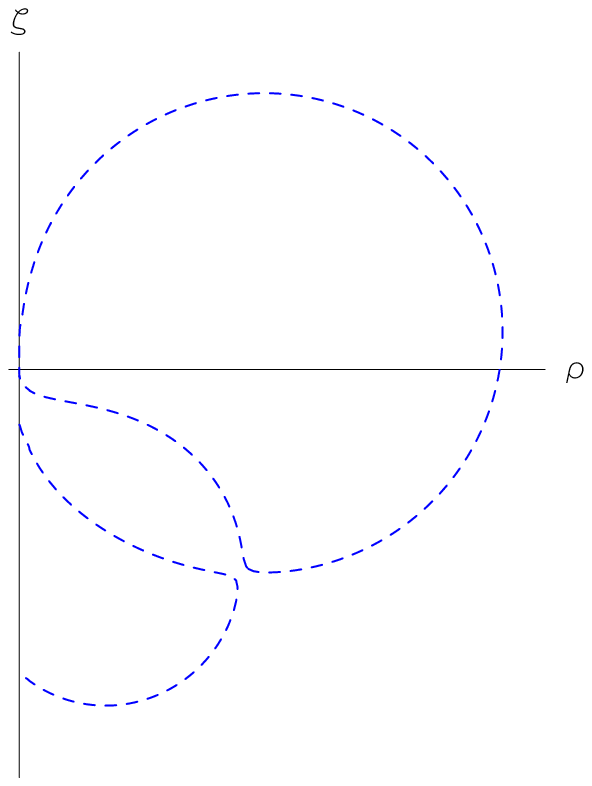}
\includegraphics[height=2in,width=.3\textwidth]{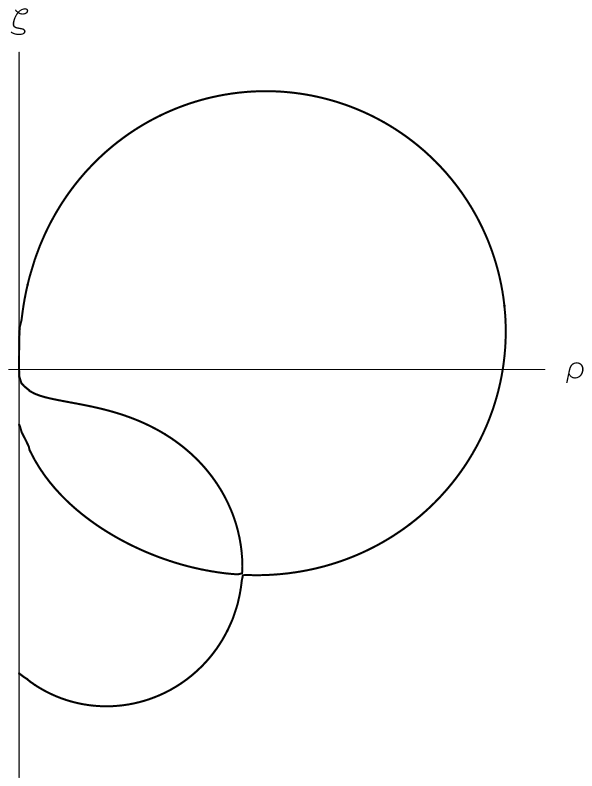}
\includegraphics[height=2in,width=.3\textwidth]{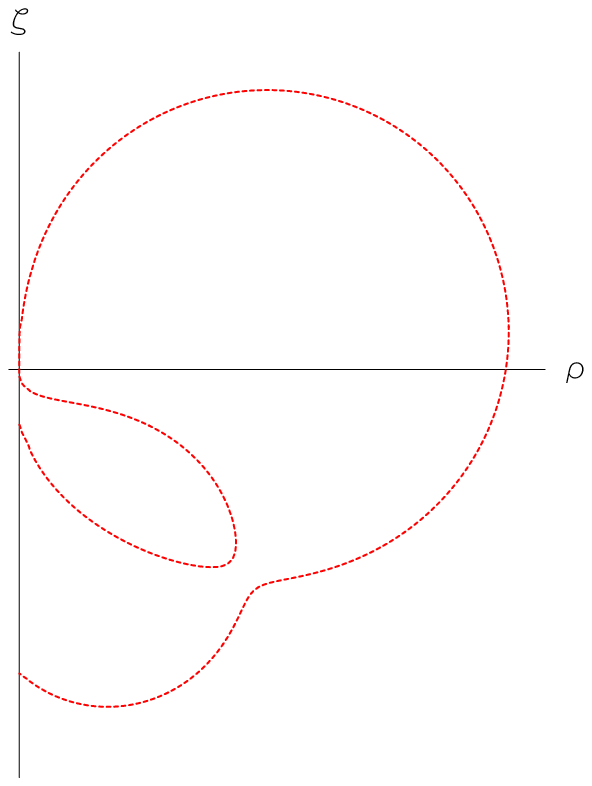}
\includegraphics[height=2in,width=.3\textwidth]{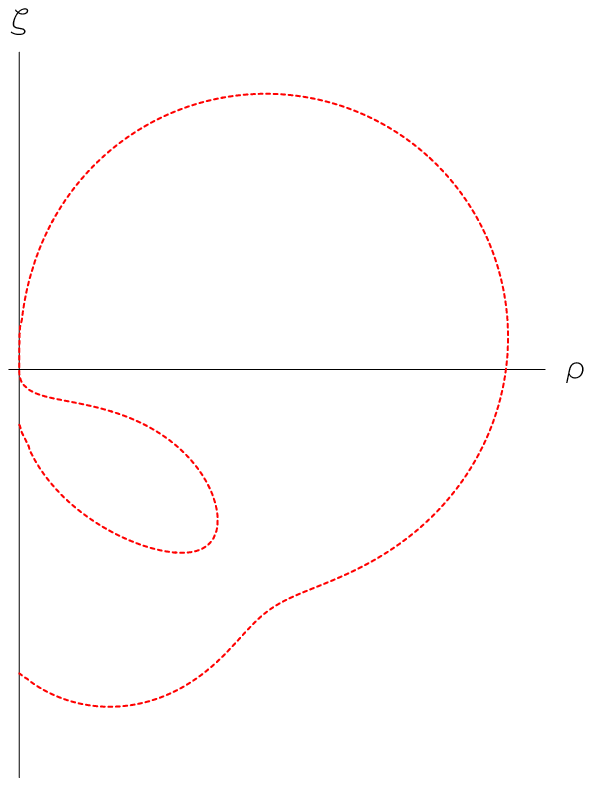}
\end{center}
\caption[]{Coalescing ergosurfaces in the ``double-Kerr" family of
metrics in the $(\varrho,\zeta)$ half-plane, with one black hole
extreme. We use a parameterization as in~\cite{Rueda:2005km}. In all
five cases the event horizon of the degenerate black hole lies at
the origin, thus $\alpha_1=0$, while $\alpha_2=-1/6$, $\alpha_3=-1$,
and with the $\beta_a$'s, $a=1,2$ of the form $\beta_a=-b_a(1+\ii)$,
where: 1) $b_1=0.6$, $b_2=1.5$;  2) $b_1=0.62$, $b_2=1.66$; 3)
$b_1=0.6218704381$, $b_2=1.668809562$; 4) $b_1=0.62$, $b_2=1.68$; 5)
$b_1=0.6$, $b_2=1.7$. Those solutions have both singular struts at
the axis and singular rings away from the region where the
coalescing of ergosurfaces occurs, but those singularities are
irrelevant for the proof that there are no local obstructions to
zeros of higher order. } \label{Coalesc}
\end{figure}
 While the value of the parameters found
numerically, for which $f$ has a zero of order two, is only
approximate, the existence of a nearby value with a second order
zero follows from what has been said above together with the
remaining results in this paper.

Similarly three ergosurfaces merging simultaneously will lead to a
zero of order precisely three, and so on.

\begin{figure}\begin{center}
\includegraphics[height=3in,width=.47\textwidth]{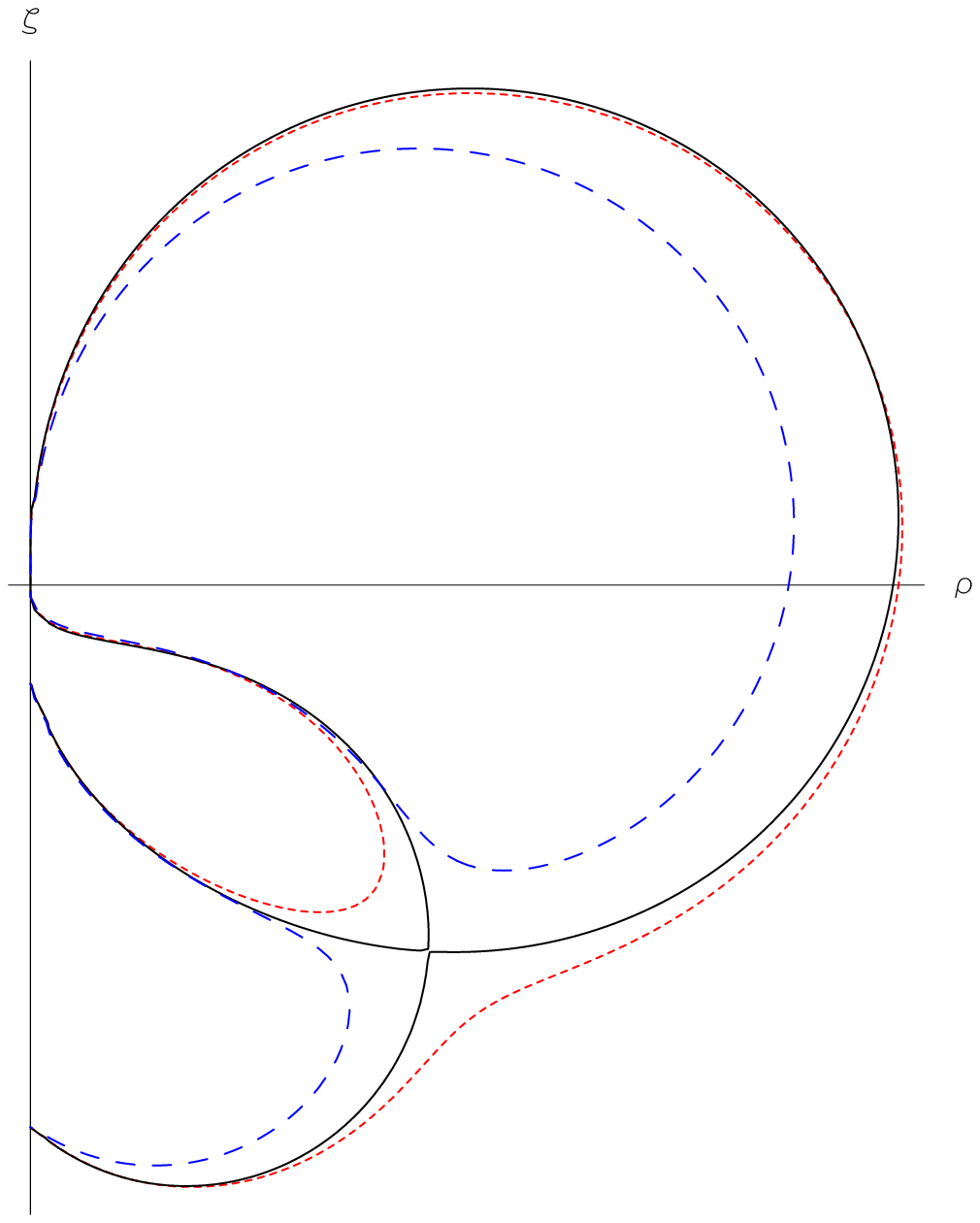}
\end{center}
\caption[]{The first, third and fifth ergosurfaces of
Figure~\ref{Coalesc} superimposed.} \label{Coalescx}
\end{figure}
\begin{figure}[ht]\begin{center}
\includegraphics[height=2in]{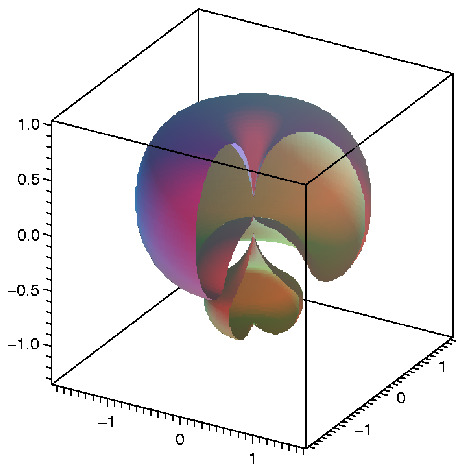}
\includegraphics[height=2in]{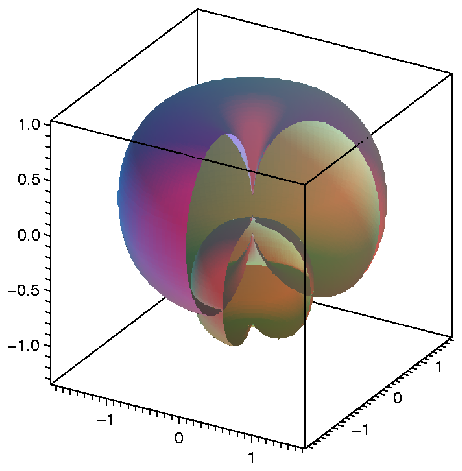}
\includegraphics[height=2in]{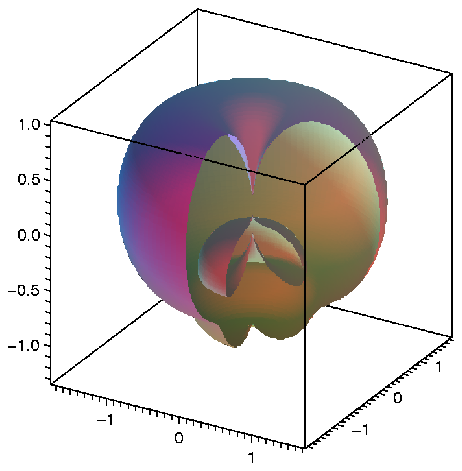}
\end{center}
\caption[]{Coordinate representation in $\R^3$ of the first, third
and fifth ergosurfaces from Figure~\ref{Coalesc}. The cusps where
the ergosurfaces meet the symmetry axis have a geometric character
and also arise in the Kerr solution~\cite{LakeKerr}.}
\label{Coalesc2}
\end{figure}

In order to study the zeros of higher order it is convenient to
consider Taylor expansions of $f$ and $b$ to order $n\ge k$,
\beal{Tayexp1}
f(\rho,\zeta) & = & \sum_{0\le i+j\le n} \frac{\zf_{i,j}}{i!j!}
(\rho-\rho_0)^i (\zeta-\zeta_0)^j+r_{n}\;,
\eea
where $$\zf_{i,j}:=\frac {\partial ^{i+j}f}{\partial^i \rho
 \partial^j\zeta}(\rho_0,\zeta_0)\;.
$$
Similarly we denote the Taylor coefficients of $b$ by $\zb_{i,j}$.

Suppose that $f$ has a zero of order $k$ at $\vec x_0$. Note that
the value $b(\rho_0,\zeta_0)=\Im \mcE(\rho_0,\zeta_0)$ is
irrelevant both for the equations and for the metric, so without
loss of generality we may assume that $\mcE(\rho_0,\zeta_0)=0$.
>From now on we assume that this is the case. Let $\mcE_k$ be the
homogeneous polynomial in $\rho-\rho_0$ and $\zeta-\zeta_0$, of
order $k$, obtained by keeping in the Taylor expansion only the
terms of first non-vanishing order, similarly $f_k$. Thus,  $f_k$
is a homogeneous polynomial in $\rho-\rho_0$ and $\zeta-\zeta_0$,
of order $k$:
\beal{Tayexp2}
f_k(\rho,\zeta) & = & \sum_{ i+j=k} \frac{\zf_{i,j}}{i!j!}
(\rho-\rho_0)^i (\zeta-\zeta_0)^j\;.
 \eea
The polynomial $f_k$ can be written in a convenient form, \eq{fkp}
below, as follows: suppose, for the moment, that $\rho-\rho_0$ is
strictly positive, by homogeneity we can then write
\bel{fkrewr}
f_k(\rho,\zeta)= (\rho-\rho_0)^k P_k(w)\;,\qquad \mbox{where}\
w:= \frac{\zeta-\zeta_0}{\rho-\rho_0}\;,
 \ee
for some non-trivial polynomial $P_k$ of order smaller than or
equal to $ k$. Let $\alpha_i\in \C$, $i=1,\ldots,n$, be distinct
zeros of  $P_k$, with multiplicities $m_i$, thus
$P_k(w)=C\,\Pi_{i=1}^n (w-\alpha_i)^{m_i}$, for some constant
$C\in \C^*$. Hence
\bel{fkp}
f_k(\rho,\zeta)= C(\rho-\rho_0)^{m_0} \,\Pi_{i=1}^n
\Big(\zeta-\zeta_0-\alpha_i(\rho-\rho_0)\Big)^{m_i}\;,
 \ee
where $m_0=k-m_1-\cdots m_n$. It should be clear that \eq{fkp}
remains true for all $\rho$, and not only for $\rho>\rho_0$ as
assumed so far.

We will need the following:

\begin{Proposition}
\label{Pbz} Assume that $f$ and $b$ are smooth near $\vec x_0$.
Then the function $b$, normalised so that $b(\vec x_0)=0$, has a
zero at $\vec x_0$ of \emph{precisely} the same order as $f$.
\end{Proposition}

\proof Let $k\in \N$ denote the order of the zero. For $k=1$ the
result  has already been established in Section~\ref{STm}, so we
assume $k\ge 2$. We then have $f=O(|\vec x - \vec x_0|^k)$,
$|Df|=O(|\vec x - \vec x_0|^{k-1})$, $\Delta f=O(|\vec x - \vec
x_0|^{k-2})$, and \eq{2.11na} shows that
$$
|Db|^2 = O(|\vec x - \vec x_0|^{2k-2})\;.
$$
Integrating radially around $\vec x_0$ gives $b=O(|\vec x - \vec
x_0|^k)$, hence the order of the zero of $b$ is larger  than or
equal to $k$.

To show the reverse inequality, suppose that $b = O(|\vec x - \vec
x_0|^{k+1})$. Inserting the Taylor expansion of $f$ into
\eq{2.11a} one finds that $f_k$ solves the equation
\bel{2.11fk}
f_k\left(\frac{\partial^2 f_k}{\partial \rho^2} + \frac{
\partial^2 f_k}{\partial \zeta^2}\right)
 =
 \left (\frac{\partial f_k}{\partial
\rho}\right)^2 + \left(\frac{
\partial f_k}{\partial \zeta}
 \right)^2
 \;.
 \ee
On the set $\Omega_k:=\{f_k>0\}$ define $u_k=\ln f_k$. Without
loss of generality, changing $f$ to $-f$ if necessary, we can
assume that $\Omega_k$ is non-empty, with $\vec x_0$ lying in the
closure of $\Omega_k$. On $\Omega_k$, \Eq{2.11fk} is simply the
statement that $u_k$ is harmonic in the variables $(\rho,\zeta)$:
\bel{uke}
\Delta_2u_k:=\frac{\partial^2 u_k}{\partial \rho^2} + \frac{
\partial^2 u_k}{\partial \zeta^2} = 0\;.
 \ee
>From \eq{fkp} we have, assuming $C\ne 0$,
$$
u_k= \ln C + m_0 \ln (\rho-\rho_0) + \sum_{i=1}^n
 m_i \ln\Big(\zeta-\zeta_0-\alpha_i(\rho-\rho_0)\Big)
 \;.
$$
Inserting into \eq{uke} one obtains
\beaa
\Delta_2 u_k & = &
 -m_0 \frac 1 {(\rho-\rho_o)^2}
 -\sum_{i=1}^n
 m_i (1+\alpha_i^2)\frac 1
 {\Big(\zeta-\zeta_0-\alpha_i(\rho-\rho_0)\Big)^2} = 0
 \;.
\eeaa
Recalling that the $\alpha_i$'s are distinct, this is only
possible if
$$
m_0=0=m_i(1+\alpha_i^2) \quad \forall \ i \;.
$$
Reordering the $m_i$'s if necessary, as $u_k$ is real-valued we
have proved that
$$
u_k = \ln C + m_1 \ln \Big((\zeta-\zeta_0)^2+(\rho-\rho_0)^2\Big)
 \quad
 \Longleftrightarrow
 \quad
 f_k = C \Big((\zeta-\zeta_0)^2+(\rho-\rho_0)^2\Big)^{m_1}\;.
$$
Subsequently,
\bel{fqm1}
f= C \Big((\zeta-\zeta_0)^2+(\rho-\rho_0)^2\Big)^{m_1} + O(|\vec x
- \vec x_o|^{k+1})\;.
 \ee
As the order of $\vec x_0$ is even, this proves
Proposition~\ref{Pbz} for all $k$ odd.

To continue, we note the following

\begin{Lemma}
\label{Lnosec} Under the hypotheses of Proposition~\ref{Pbz}, let
$\vec x_0$ be a zero of order two. Then the quadratic form defined
by the Hessian $DDf(\vec x_0)$ of $f$ has signature $(+-)$ or
$(-+)$. This implies that second order zeros of $f$ are
isolated.
\end{Lemma}

\begin{Remark}
\label{Rnosec} For further use we note that the derivation of
\eq{firsset}-\eq{secset} only uses the truncated equations
\eq{2.11fkp}-\eq{2.11bkp} below. Furthermore, the calculations here
--- and therefore their conclusions --- remain valid when a
supplementary error term $o(|\vec x - \vec x_0|^3)$ is allowed at
the right-hand-side of \eq{2.11na}.
\end{Remark}

\proof
 The result is obtained by a calculation, the simplest way proceeds as in the proof of Theorem~\ref{T1} below. Alternatively,
one can use {\sc Maple} or {\sc Mathematica}, the interested reader
can download the worksheets
from~\url{http://th.if.uj.edu.pl/~szybka/CMS}; that last calculation
has been done as follows: Consider the polynomials $W_a$, $a=1,2$,
obtained by inserting the Taylor expansion of $f$ and $b$, with
$\zf=D\zf=0$, into equations obtained by multiplying \eq{2.11na} and
\eq{2.11nb} with $\rho$. The requirement that those polynomials
vanish up-to-and-including order two imposes the following
alternative sets of conditions:
\bea
\mbox{I.} & \zb_{2,0}=\zb_{1,1}=\zb_{0,2}=\zf_{1,1}=0\;,\quad
\zf_{2,0}=\zf_{0,2}\in \R\;,
 \label{firsset}
\\
\mbox{II.} &  \zf_{2,0}=-\zf_{0,2}=-\zb_{1,1}\in \R\;,\quad
\zb_{0,2}=-\zf_{1,1}=-\zb_{2,0}\in \R\;,
 \eeal{secset}
as well as a set which is related to II.  above by exchanging $b$
with $-b$. The first set leads to $\zf_{0,2}=0$ when requiring that
the polynomials $W_a$ just defined vanish to one order higher, so
that the first set cannot occur for zeros of second order. One then
checks that the set II. leads to Lorentzian signature of $DDf$,
unless vanishing. \qed

Clearly the Hessian of $f$ given by \eq{fqm1} does not have
indefinite signature when $m_1=1$, proving Proposition~\ref{Pbz}
for zeros of order two.

It remains to consider $m_1\in \N$ satisfying $m_1\ge 2$. Replacing
$f$ by $-f$ if necessary,
it follows from \eq{fqm1} that $f$ is strictly positive in a
neighborhood of $\vec x_0$, so that we can define
$$
g:= f^{1/m_1}\;.
$$
Usual arguments (cf., e.g.,~\cite{Malgrange2}), show that $g$ is
smooth and has a zero of order two at $\vec x_0$. From \eq{2.11na}
one has
\beal{2.11nax}
&
 \displaystyle
  g \Delta g - |Dg|^2=\frac 1 {m_1}g^{2}\underbrace{\frac{|Db|^2}{f^2}}_{O(1)}=O(|\vec x- \vec x_0|^4)\;.
 &
 \eea 
Taylor expanding $g$ up to order $o(|\vec x- \vec x_0|^4)$ and
inserting into \eq{2.11nax} gives $C=0$ (see Remark~\ref{Rnosec}),
proving Proposition~\ref{Pbz}.
 \qed

\subsection{Simple zeros}\label{SSz}

A zero of $f$ of order $k$ will be said to be \emph{simple} if all
the $\alpha_i$'s in \eq{fkp} are real and have multiplicities one,
with $m_0\in \{0,1\}$.  We will show below that zeros of finite
order of solutions of Ernst equations are simple. Somewhat to our
surprise, for such zeros Theorem~\ref{Tmain} generalises as follows:

\begin{Theorem}
\label{Tmain2}
 The conclusions of Theorem~\ref{TA} are valid under
the supplementary condition that $f$ has only simple zeros at the
$\mcE$--ergosurface $E_f:=\{f=0,\rho>0\}$.
\end{Theorem}

\proof As pointed out by Malgrange~\cite[end of
Section~3]{Malgrange2}, simplicity implies that near $\vec x_0$
there exist smooth functions $\phi_a$, $a=1, \ldots, k$, with
$\phi_a(\vec x_0)=0$ and with nowhere-vanishing gradient, together
with a strictly positive smooth function $g$ such that we can
write
\bel{ffact}
f= \phi_1\cdots \phi_k g
 \;.
\ee
(Supposing that $m_0=0$, the $\phi_a$'s have the Taylor expansion
$\phi_a= \zeta-\zeta_0 - \alpha_a (\rho-\rho_0)+O(|\vec x-\vec
x_0|^2)$; if $m_0=1$, then one has $\phi_1=\rho-\rho_0+O(|\vec
x-\vec x_0|^2)$, with the remaining Taylor expansions of the same
form as before. For $k=2$ this is a special case of Morse's
theorem~\cite[Theorem 6.9, p. 65]
{GolubitskyGuillemin}.)

\Eq{ffact} shows that $E_f$ is, near $\vec x_0$, the union of the
smooth submanifolds $\{\phi_a=0\}$. On each of those $Df$ is
non-vanishing, except at the origin. Passing to a small neighborhood
of $\vec x_0$ if necessary, we can assume that each of the sets
$\{\phi_a=0,Df\ne 0\}$ has precisely two components.

Consider a connected component of $\{\phi_1=0\}$, by
Section~\ref{STm} \Eq{3.1} holds there. Suppose that the lower sign
arises on this component, then the same lower sign has to arise on
the remaining component of $\{\phi_1=0\}$, because the inversion
$\vec x-\vec x_0 \to -\vec x + \vec x_0$ maps each component to the
accompanying one up to quadratic terms, and because $Db$ has, in the
leading order of its Taylor development, the same parity as $Df$ by
Proposition~\ref{Pbz}.

We consider the function $\sigma_\rho$ as in \eq{eq1n}, an identical
argument applies to $\sigma_\zeta$ and to $\kappa_\rho$,
$\kappa_\zeta$. Using a coordinate system $(y^1,y^2)$ with
$\phi_1=y^1$ we have a Taylor expansion
\bel{Tsiex} \sigma_\rho(y^1,y^2) = \sigma_{\rho,0}(y^2) +
\sigma_{\rho,1}(y^2)y^1 + \sigma_{\rho,2}(y^1,y^2)(y^1)^2\;.
 \ee
Note that $f$ has a simple zero away from the origin on the axis
$\{y^1=0\}$, so by the results in Section~\ref{STm} the functions
$\sigma_{\rho,0}$ and $\sigma_{\rho,1}$ vanish there. By
continuity they also vanish at the origin, thus $\sigma_\rho$
factorises as
$$
\sigma_\rho = \sigma'_{\rho} \phi_1^2
$$
 for a smooth function
$\sigma'_{\rho}:=\sigma_{\rho,2}$.

We introduce a new coordinate system $(z^1,z^2)$ in which
$z^1=\phi_2$. We Taylor expand $\sigma'_\rho$ as in \eq{Tsiex},
with the $y^i$'s there replaced by $z^i$'s, etc. The equations
\beaa
 0&=&\sigma_\rho|_{z^1=0,z^2\ne 0}= \sigma'_\rho|_{z^1=0,z^2\ne 0}
\underbrace{\phi_1^2|_{z^1=0,z^2\ne 0}}_{\ne 0}\;,
 \\
 0&=&d\sigma_\rho|_{z^1=0,z^2\ne 0}= d\sigma'_\rho|_{z^1=0,z^2\ne 0}
\underbrace{\phi_1^2|_{z^1=0,z^2\ne 0}}_{\ne 0}
 \\
 && +
2\underbrace{\sigma'_\rho|_{z^1=0,z^2\ne 0}}_{= 0} (\phi_1
d\phi_1)|_{z^1=0,z^2\ne 0}\;, \eeaa
show that the function $\sigma'_\rho$ vanishes, together with its
first derivatives, away from the origin on the axis $\{z^1=0\}$.
We conclude as before that $\sigma'_\rho$ factorises as
$\sigma'_\rho = \sigma''_{\rho} \phi_2^2$ for a smooth function
$\sigma''_{\rho}$, hence $\sigma_\rho$ factorises as
$$
\sigma_\rho = \sigma''_{\rho} \phi_1^2\phi_2^2\;.
$$
Continuing in this way, in a finite number of steps one obtains
$$
\sigma_\rho = \hat \sigma_{\rho} \phi_1^2\cdots \phi_k^2\;,
$$
and the result easily follows.
 \qed

\subsection{Zeros of finite order are simple}
\label{Szos}

Consider a zero of $f$ of order $k<\infty$, with $\rho_0>0$, then
the leading order Taylor polynomials $f_k$ and $b_k$ solve the
truncated equations
\bel{2.11fkp}
f_k\left(\frac{\partial^2 f_k}{\partial \rho^2} + \frac{
\partial^2 f_k}{\partial \zeta^2}\right)
 =
 \left (\frac{\partial f_k}{\partial
\rho}\right)^2 + \left(\frac{
\partial f_k}{\partial \zeta}
 \right)^2
- \left (\frac{\partial b_k}{\partial \rho}\right)^2 -\left(\frac{
\partial b_k}{\partial \zeta}
 \right)^2
 \;,
 \ee
 \bel{2.11bkp}
f_k\left(\frac{\partial^2 b_k}{\partial \rho^2} + \frac{
\partial^2 b_k}{\partial \zeta^2}\right)
 =
  2\Big(\frac{\partial f_k}{\partial
\rho}\frac{\partial b_k}{\partial \rho} +
 \frac{
\partial f_k}{\partial \zeta}\frac{
\partial b_k}{\partial \zeta}\Big)
 \;.
 \ee

Let
\bel{Ek} f_k+\ii b_k\equiv \mcE_k = \alpha (z-z_0)^k\;,
 \ee
where $\alpha\in \C$, with $z=\rho+\ii \zeta$. It is
straightforward, using the Cauchy-Riemann equations, to check that
functions of this form satisfy \eq{2.11fkp}-\eq{2.11bkp}, for all
$k\in \N$. (In fact, both the left- and right-hand-sides then vanish
identically.) Those solutions have been found by inspection of the
solutions found by {\sc Maple} for $k=2$ and by {\sc
Singular}~\cite{Singular,Greuel} for $k=3$ and $4$. In fact, both
the {\sc Singular}--generated solutions, as well as our remaining
computer experiments using {\sc Singular}, played a decisive role in
our solution of the problem at hand.

 Let us show that:

 \begin{Lemma}
 \label{Ls} Zeros of $f_k$ given by \eq{Ek} are
{simple}.
 \end{Lemma}

 \proof Indeed, the equation $f_k=0$ is equivalent to
$$
\alpha (z-z_0)^k = \ii \beta\;,
$$
for some $\beta \in \R$.
 This is easily solved; we write $\alpha= |\alpha| e^{\ii \theta}$, and set
$$
\alpha_\ell=\tan\Big(\frac{{(2 \ell +1)\pi} -2\theta } {2k} \Big)\;,
\quad  \ell=1,\ldots,k\;.
$$
%
Assuming $\alpha_\ell\ne \pm \infty$ for all $\ell$,  we obtain  $k$
\emph{distinct} real lines $z_0 + \R (1+\ii \alpha_\ell)$ on which
$\Re\mcE_k$ vanishes, and simplicity follows. The remaining cases
are analysed similarly, and are left to the reader. \qed

  Another non-trivial, ``polarised", family of
solutions of \eq{2.11fkp}-\eq{2.11bkp} is provided by $b_k=0$,
$f_k=C\Big((\rho-\rho_0)^2+(\zeta-\zeta_0)^2\Big){}^m$, $m\in \N$.
As mentioned in Section~\ref{Sns}, there exist associated
\emph{static} solutions of the Ernst equations. However, as already
pointed out (compare Remark~\ref{Rnosec}), neither those, nor any
other solutions with this $f_k,b_k$, are smooth across $E_f$.

Setting $z=\rho-\rho_0+\ii (\zeta-\zeta_0)$,
the equations satisfied by $\mcE_k=f_k+\ii b_k$ take the form%
\newcommand{\oz}{{\bar z}}%
\bel{comEr} (\mcE_k + \overline {\mcE}_k)\frac{\partial^2
\mcE_k}{\partial z\partial\oz} = 2\frac{\partial \mcE_k}{\partial
z}\frac{\partial \mcE_k}{\partial\oz} \;. \ee

Since $\mcE_k$ is a polyhomogeneous polynomial in $x$ and $y$, it
can be written as
$$
\mcE_k = \sum_{m=0}^k \beta_m z^m \oz^{k-m} \;.
$$
Inserting this into \eq{comEr} we obtain
$$
\sum_{1\le m+j\le 2k-1}m\beta_m\{(2j-k-m)\beta_j
+(k-m)\bar\beta_{k-j}\}z^{m+j-1}\oz^{2k-m-j-1}=0 \;.
$$
Hence, for $1\le \ell \le 2k-1$,
\bel{newmain}
\sum_{m+j=\ell}\{(k-m)\bar{\beta}_{k-j}-(k+m-2j)\beta_j \}m\beta_m=0
\;. \ee
Since $\ell=0$ is trivial, we obtain $2k-2$ equations for $k+1$
numbers $\beta_m$, which should be rather restrictive, especially
for $k\ge 3$. Nevertheless, as already pointed out, there exist
non-trivial solutions. It is instructive to find them directly by
inspection of \eq{newmain}. First, there is the obvious solution
$\beta_m=0$ for $m\ge 1$, which corresponds to an anti-holomorphic
$\mcE_k= \beta_0 \oz^k$. Next, one checks that a collection with
$\beta_k\ne 0$ but $\beta_m=0$ for $m< k$ provides a solution, which
corresponds to a holomorphic $\mcE_k= \beta_k z^k$. Finally, when
$k=2n$, one checks that $\beta_n\in \R$, but $\beta_m=0$ for $m\ne
k/2$,
 is a solution, which
corresponds to a real $\mcE_{2n}= \beta_n z^n\oz^n=
\beta_n(x^2+y^2)^n$.

The computer algebra program {\sc Singular} can be used to show that
the above exhaust the list of solutions for $k$ less than or equal
to eight\footnote{The {\sc Singular} input file is available on
URL~\url{http://th.if.uj.edu.pl/~szybka/CGMS}}.  This turns out to
be true for all $k<\infty$:

\begin{Theorem}
\label{T1}
  These are all solutions: thus the homogeneous polynomial $\mcE_k$ is either
  holomorphic, or anti-holomorphic, or real and radial.
\end{Theorem}

%

\proof The case $k=1$ is a straightforward calculation, so we assume
$k>1$.

If $\mcE_k$ is a solution, then so is its complex conjugate; this
implies that if an ordered collection $\{\beta_m\}_{0\le m\le k}$
satisfies \eq{newmain}, then so does $\{\bar\beta_{k-m}\}_{0\le m\le
k}$. Inserting this into \eq{newmain} one obtains, again for $1\le
\ell \le 2k-1$,
\bel{cc1} \sum_{m+j=\ell}\{(k-m){\beta}_{j}-(k+m-2j)\bar\beta_{k-j}
\}m\bar\beta_{k-m}=0 \;. \ee
Consider \eq{newmain} with $\ell=1$; since $1\le m\le \ell$ this
enforces $m=1$, $j=0$, giving
\bel{newmainx} \{(k-1)\bar{\beta}_{k}-(k+1)\beta_0 \}\beta_1=0 \;.
\ee
Similarly \eq{cc1} with $\ell=1$ gives
\bel{newmain2} {\{(k-1){\beta}_{0}-(k+1)\bar\beta_k
\}}\bar\beta_{k-1}=0 \;. \ee

Suppose, first, that $\beta_0\ne 0$. We will use  induction
arguments to establish the implication \eq{imp1} below. So assume,
for contradiction, that
 $\beta_1\ne 0$. Then $\bar
\beta_k=(k+1)\beta_0/(k-1)\ne 0$ from \eq{newmainx}, inserting into
\eq{newmain2} we obtain  $\beta_{k-1}=0$. But then \eq{cc1} with
$\ell=2$ gives
$$
0=\{(k-2){\beta}_{0}-(k+2)\bar\beta_{k}
\}\bar\beta_{k-2}=\underbrace{\frac{(k-2)(k-1)-(k+2)(k+1)}{k-1}\beta_{0}
}_{\ne 0}\bar\beta_{k-2}\;,
 $$
hence $\beta_{k-2}=0$. \Eq{cc1} with $\ell=3$ similarly gives now
$\beta_{k-3}=0$. Continuing in this way one concludes in a finite
number of steps that $\beta_1=0$, a contradiction. It follows that
$\beta_0\ne 0$ enforces $\beta_1= 0$.

Assume now, again for contradiction, that $\beta_0\ne 0$ and
$\beta_1=0$ but $\beta_2\ne 0$. \Eq{newmain} with $\ell=2$ gives
$$
\{(k-2)\bar{\beta}_{k}-(k+2)\beta_0 \}\beta_2=0 \;.
 $$
If $k=2$ we obtain immediately a contradiction; otherwise $\bar
\beta_k=(k+2)\beta_0/(k-2)\ne 0$, inserting into \eq{newmain2} we
find  $\beta_{k-1}=0$. But then \eq{cc1} with $\ell=2$ gives
$$
0=\{(k-2){\beta}_{0}-(k+2)\bar\beta_{k}
\}\bar\beta_{k-2}=\underbrace{\frac{(k-2)^2-(k+2)^2}{k-1}\bar\beta_{k}
}_{\ne 0}\bar\beta_{k-2}\;,
 $$
hence $\beta_{k-2}=0$.  Continuing in this way one concludes in a
finite number of steps that $\beta_2=0$, a contradiction. This shows
that  $\beta_0\ne 0$ and $\beta_1=0$ but $\beta_2\ne 0$ is
incompatible with the equations.

It should be clear to the reader how to iterate this argument to
obtain the implication
\bel{imp1}
 \beta_0\ne 0 \ \mbox{ implies } \ \beta_m=0 \ \mbox{ for } \  m=1,\ldots,k\;.
\ee

Using symmetry under complex conjugation, the hypothesis $\beta_k\ne
0$ leads to $\beta_m=0$ for $m=0,\ldots,k-1$.

 It remains to analyse
what happens when $\beta_0=\beta_k=0$, which we assume from now on.
Suppose (for contradiction if $k>2$) that $\beta_1\ne 0$. Recalling
that $k>1$, \eq{newmain}-\eq{cc1} with $\ell=2$ give
 $$
(\bar{\beta}_{k-1}-\beta_1 )\beta_1=0 = ({\beta}_{1}-\bar\beta_{k-1}
)\bar\beta_{k-1}\;.
 $$
If $k=2$ we obtain $\beta_1\in \R$, and we are done.

Otherwise
 $\beta_{k-1}=\bar \beta_1 \ne 0$ and
\eq{cc1} with $\ell=3$ gives
$$
\{(k-1)\beta_2-(k+1)\bar\beta_{k-2}\}\beta_1=0\;.
$$
When $k=3$ this gives a contradiction,  and the result is
established for this value of $k$.


For $k\ge 4$ the proof will be finished by more induction arguments,
as follows: Suppose, to start with, that $\beta_k=0$ and that there
exist $k_0,k_1\in \N$, $1\le k_1\le k_0\le k/2$, such that
$\beta_m=0$ for $0\le m\le k_0-1$ and for $k-k_1< m\le k$
 but
$\beta_{k_0}\ne 0$. (The case $k_1
>k_0$ can be reduced to this one by replacing $\mcE_k$ with its
complex conjugate.) With these hypotheses \eq{newmain} can be
rewritten as
\bel{d1} \sum_{k_0\le m \le
\min(k-k_1,\ell-k_1)}\{(k-m)\bar{\beta}_{k-(\ell-m)}-(k+3m-2\ell)\beta_{\ell-m}
\}m\beta_m=0 \;. \ee
\Eq{d1} with $\ell=k_0+k_1\le k$ gives:
$$
(k-k_0)\bar{\beta}_{k-k_1}=(k+k_0-2k_1)\beta_{k_1}
 $$
which equals zero unless $k_1=k_0$. It follows that we  can
without loss of generality assume that  $k_1=k_0$ and
$$
{\beta}_{k-k_0}=\bar\beta_{k_0}
 \;.
 $$
 We can now rewrite \eq{cc1}
as
\bel{d2} \sum_{k_0\le m \le
\min(k-k_0,\ell-k_0)}\{(k-m){\beta}_{\ell-m}-(k+3m-2\ell)\bar\beta_{k-(\ell-m)}
\}m\bar\beta_{k-m}=0 \;. \ee
Suppose that $k=2k_0$; then \eq{d2} leads immediately to the
restriction $\beta_{k_0} \in \R$, giving a real radial solution, as
desired.
Otherwise, choosing $\ell=2k_0+1$ in \eq{d2} one obtains
$$
(k-k_0)k_0\beta_{k_0+1}= [(k-k_0)k_0+2] \bar \beta_{k-k_0-1}\;.
$$
 \Eq{d1} with $\ell=2k_0+1$ gives
$$
(k-k_0)k_0\bar\beta_{k-k_0-1}= [(k-k_0)k_0+2] \beta_{k_0+1}\;.
$$
It follows that
$$
\beta_{k_0+1}=  \beta_{k-k_0-1}=0\;.
$$

Our aim now is to show \eq{verygood} below, by a last induction. So,
suppose there exists $k_2\in \N$ satisfying $k_0< k_2<k-k_0$ such
that $\beta_m=0$ for $k_0<m<k_2$ and for $k-k_2<m<k-k_0$; we have
shown that this is true with $k_2=k_0+2$.  \Eq{d2} with
$\ell=k_0+k_2$ gives
$$
(k-k_0)k_0\beta_{k_2}= [(k-k_0)k_0+2(k_2-k_0)^2] \bar
\beta_{k-k_2}\;.
$$
But from \eq{d1} again with $\ell=k_0+k_2$ one obtains
$$
(k-k_0)k_0\bar\beta_{k-k_2}= [(k-k_0)k_0+2(k_2-k_0)^2]
\beta_{k_2}\;.
$$
This allows us to conclude that
\bel{verygood}
\beta_m=0 \ \mbox{ except if } \ m=k_0 \ \mbox{ or if } \
m=k-k_0\;, \ \mbox{ with } \ \beta_{k-k_0}=\bar \beta_{k_0}
 \;.
 \ee
\Eq{d2} with $\ell=k$ gives now $\beta_{k_0}=0$ (recall that we have
assumed $k\ne 2k_0$), a contradiction, and the theorem is proved.
 \qed

We can now pass to the

\medskip

\noindent {\sc Proof of Theorem~\protect\ref{TA}}: Theorem~\ref{T1}
gives the list of all possible $\mcE_k$'s. The real ones do not lead
to smooth $f$'s by Proposition~\ref{Pbz}. The holomorphic  ones lead
to simple zeros by Lemma~\ref{Ls};
the same is true for the anti-holomorphic ones, because the
condition of simplicity is preserved by complex-conjugation of
$\mcE$. The result follows now from Theorem~\ref{Tmain2}. \qed

\bigskip

\noindent{\sc Acknowledgements} We wish to thank G.~Alekseev,
M.~Ansorg, R.~Beig, M.-F. Bidaut-V\'eron, M.~Brickenstein, S.~Janeczko,
J.~Kijowski, G.~Neugebauer, D.~Petroff and L.~V\'eron for useful
comments or discussions.
We acknowledge hospitality and financial support from the Newton
Institute, Cambridge (PTC, RM, SSz), as well as the AEI, Golm
(PTC) during work on this paper.

\bigskip

\bibliographystyle{amsplain}
\bibliography{
../../references/reffile,%
../../references/newbiblio,%
../../references/bibl,%
../../references/howard,%
../../references/myGR,%
../../references/newbib,%
../../references/Energy,%
../../references/netbiblio,%
../../references/PDE}
\end{document}